\documentclass[10pt, letterpaper]{article}

\usepackage[T1]{fontenc}
\usepackage[utf8]{inputenc}
\usepackage{authblk}
\usepackage{fvextra}
\usepackage{mathtools}
\usepackage{parskip}
\usepackage{fullpage}
\usepackage{times}
\usepackage{latexsym}
\usepackage{nameref}
\usepackage{enumitem}
\usepackage{float}
\usepackage{graphicx}
\usepackage{amssymb}
\usepackage{booktabs}
\usepackage[font=small]{caption, subcaption}
\usepackage{multirow}
\usepackage{tabularx}
\usepackage{makecell}
\usepackage{hyperref}
\usepackage[autostyle=true]{csquotes}
\usepackage[numbers]{natbib}
\usepackage{xcolor, colortbl}
\usepackage{xurl}

\hypersetup{
	colorlinks=true,
	citecolor=blue,
	linkcolor=blue,
	urlcolor  = blue
}

\emergencystretch=\maxdimen
\makeatletter
\renewcommand\AB@affilsepx{\hspace{1em} \protect\Affilfont}
\makeatother

\renewcommand*{\Affilfont}{\fontsize{8.5}{10.5}\selectfont\normalfont}

\author[1]{Neil K. R. Sehgal\thanks{These authors contributed equally.}\thanks{To whom correspondence should be addressed. E-mail: \href{mailto:nsehgal@upenn.edu}{\nolinkurl{nsehgal@upenn.edu}}}}
\author[2]{Manuel Tonneau\protect\footnotemark[1]}
\author[1]{Dunigan Folk}
\author[1]{Lyle Ungar}
\author[1]{Sharath Chandra Guntuku}
\affil[1]{University of Pennsylvania}
\affil[2]{Hasso Plattner Institute, University of Potsdam}

\renewenvironment{abstract}
 {\small
  \begin{center}
  \bfseries \abstractname\vspace{-0.5em}
  \end{center}
  \list{}{%
    \setlength{\leftmargin}{6.25mm}%
    \setlength{\rightmargin}{\leftmargin}%
  }%
  \item\relax}
 {\endlist}

\newcommand{\beginsupplement}{
  \clearpage
  \captionsetup{labelsep=period}
  \renewcommand{\figurename}{Fig.}
  \setcounter{section}{0}
  \setcounter{figure}{0}
  \setcounter{table}{0}
  \renewcommand\thesection{S\arabic{section}}
  \renewcommand\thesubsection{\thesection.\arabic{subsection}}
  \renewcommand\thefigure{S\arabic{figure}}
  \renewcommand\thetable{S\arabic{table}}
}

\title{AI-Assisted Submissions in Online Research Are Rare and Highly Concentrated but Routinely Approved}
\date{}

\begin{document}

\maketitle

\vspace{-24pt}

\begin{abstract}
Online research platforms underpin much of what science claims about people, on the assumption that a human produced each response. Generative AI threatens that assumption by letting participants delegate responses to a chatbot, yet how often they do so remains unclear because prior estimates rely on self-report or automated detection rather than direct observation. We surveyed 2,500 workers on a high quality online research platform and directly observed AI assistance by linking donated ChatGPT histories to platform submission records of weekly ChatGPT users, covering 712,930 submissions across more than 127,000 studies. Although one in eight surveyed workers reported ever using AI on a study and 68\% of observed workers had done so, assistance appeared in only 1\% of submissions, with no detectable increase over more than 3 years. Assistance was often temporally localized within tasks and highly concentrated among workers, with 5\% accounting for 64\% of assisted submissions. Where assistance occurred, workers were virtually always paid, even when study instructions prohibited AI use, with overall approval rates similar to those for unassisted submissions. Half of assisted submissions involved bounded responses, outside the scope of the platform's LLM detector for open-ended responses. Taken together, our results do not support the view that AI assistance currently poses an existential threat to online research, but reveal limited payment consequences for prohibited use and gaps in platform safeguards, leaving platforms poorly prepared should that threat materialize.
\end{abstract}


\section*{Introduction}

Online research platforms serve as standard infrastructure for collecting human input at scale, from behavioral research to AI alignment. Generative AI threatens that input's integrity by producing responses a study intended to elicit from a person. While AI agents completing studies with no human involvement currently make up no more than 1\% of respondents on most platforms \cite{ref1}, the more likely threat, workers consulting a chatbot while completing a task, is already well documented \cite{ref2,ref3}.

How often AI assistance occurs, however,{} remains unclear. Published estimates range from 0\% to 35\% \cite{ref2,ref3,ref4,ref5,ref6}, but typically come from single studies and task types, sometimes selected because they invite AI use \cite{ref2}, making them difficult to compare \cite{ref7}. None is based on direct observation: estimates rely on self-report \cite{ref8} or error-prone detection tools that disagree markedly \cite{ref3}. Three questions remain: what share of submissions is affected, whether assistance concentrates among a few workers, and what its forms and payment outcomes reveal about current safeguards.

We address these questions in two studies of U.S., U.K., and Canadian workers on Prolific, one of the highest quality research recruitment platforms \cite{ref9}. We first surveyed 2,500 workers about AI assistance, with the sample broadly resembling the active Prolific pool on most measured demographics within each country. We then directly observed such assistance by linking previously donated ChatGPT histories to Prolific submission records for 408 weekly ChatGPT users whose demographic profiles were generally similar to those of survey participants. The linked records covered 712,930 submissions across more than 127,000 studies conducted from December 2022 to February 2026. We count a submission as affected when visible ChatGPT output answered or materially contributed to a response, regardless of permission.

\section*{Results}

\begin{figure}[htbp]
\centering
\includegraphics[width=\linewidth]{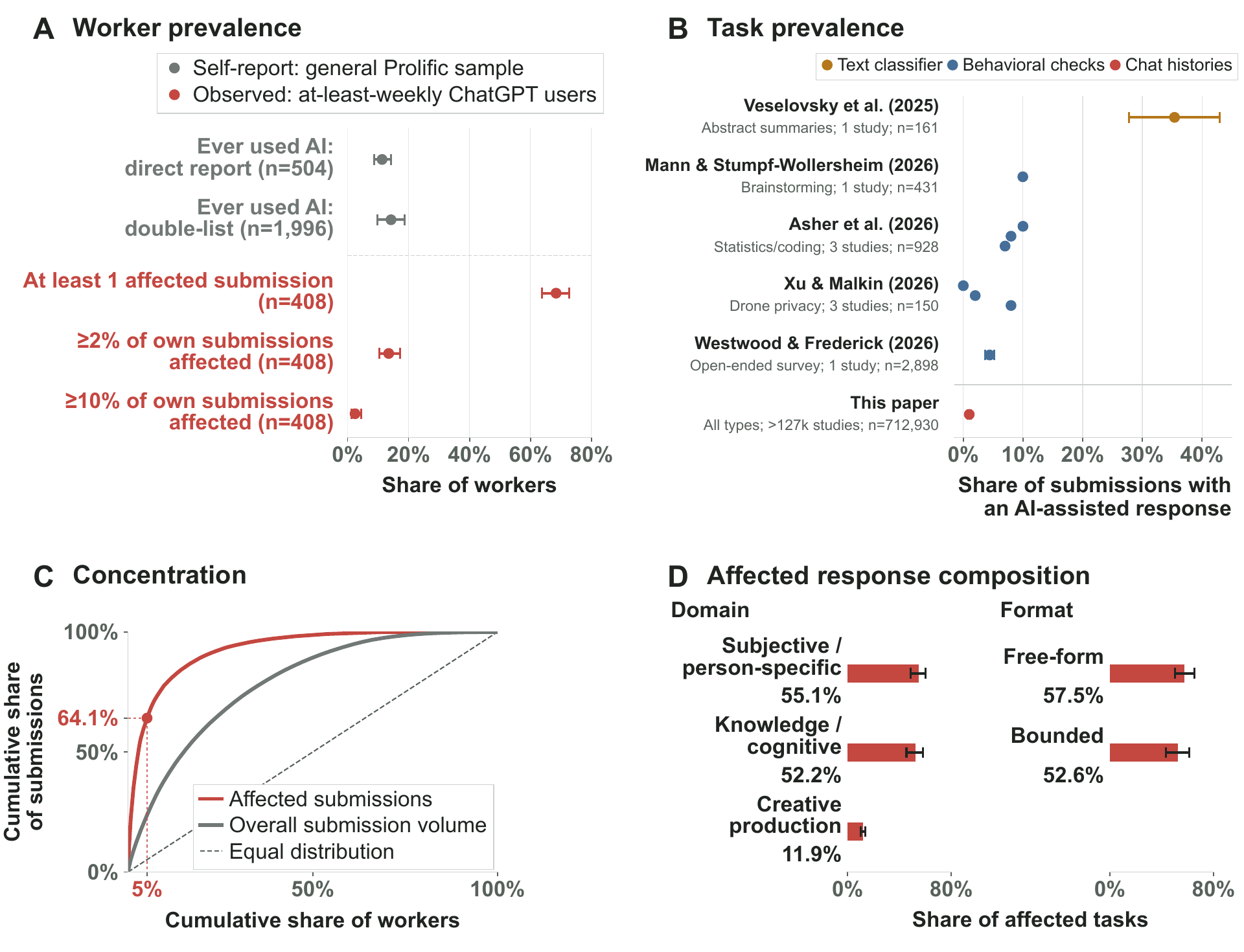}
\caption{\textbf{AI assistance is familiar among workers yet rare across submissions, concentrated in a few workers and spanning response domains and formats.}
(A) Worker-level prevalence: self-reports from a general Prolific sample (top two rows) and direct observations from at-least-weekly ChatGPT users' donated histories (bottom three). Ever-use was common among observed workers, but assistance on a substantial share of their submissions was uncommon.
(B) Submission-level prevalence versus comparable published Prolific estimates, with task type and scope shown. AI assistance appeared in 1\% of submissions, below most published estimates.
(C) Lorenz curves across 408 workers, ranked from most to fewest submissions separately for each curve. Affected submissions were more concentrated (Gini, 0.856; 95\% CI, 0.819--0.880) than overall submissions (Gini, 0.575; 95\% CI, 0.545--0.603).
(D) Response domains and formats among 7,100 affected submissions. Categories are nonexclusive because a submission could contain differently labeled exchanges.
}
\label{fig:prevalence}
\end{figure}

\textbf{Using AI on a submission is a familiar behavior among workers (Fig.~\ref{fig:prevalence}A).} We recruited 2,500 Prolific workers without screening for AI use and randomly assigned them to either a direct-question condition or an indirect list experiment that does not reveal individual answers. About 63\% of these individuals reported using ChatGPT at least weekly. Asked directly, 11.31\% reported ever using AI to help with a Prolific submission (95\% CI, 8.73\%--14.29\%). The list experiment estimated 14.24\% (9.73\%--18.70\%), a difference that was not statistically significant and provided no clear evidence that anonymity increased reporting. Direct observation of the 408 workers who donated their ChatGPT histories put the share who had ever used ChatGPT for assistance at 68.4\%. Even among them, only 13.5\% had assistance on at least 2\% of their submissions. Thus, having used AI at least once was common among these workers, while using it on a substantial share of submissions was much less common. To assess how much research data is affected, we next examine the proportion of submissions involving AI assistance.

\textbf{AI assistance appeared in only 1\% of submissions, below most published estimates (Fig.~\ref{fig:prevalence}B).} We identified assisted submissions using a validated classifier applied to temporally overlapping ChatGPT exchanges. Across the 712,930 Prolific submissions made by 408 weekly ChatGPT users while their ChatGPT histories were observable, assistance appeared in 7,100 (1.00\%, worker-cluster bootstrap 95\% CI 0.72\%--1.32\%). Assistance was also temporally localized within submissions: 40\% of affected submissions contained a single ChatGPT exchange, and among the remaining 60\%, the first and last ChatGPT messages spanned a median 29\% of study duration, suggesting workers consult ChatGPT on particular responses rather than throughout.

\textbf{Assistance was concentrated among a small share of workers (Fig.~\ref{fig:prevalence}C).} Ranked by affected-submission count, the top 5\% of workers (21 of 408) accounted for 64.1\% of affected submissions and the top 10\% for 78.2\%, far exceeding their share of overall submissions, so differences in study volume alone do not explain the concentration. Excluding the top 5\% reduced the affected submission rate from 1.00\% to 0.40\%, a 59.4\% relative reduction, so removing a small number of workers would remove most assistance, provided those workers can be identified in advance.

\textbf{Assistance reached subjective and cognitive responses alike and was not confined to open-ended text (Fig.~\ref{fig:prevalence}D).} We classified AI-assisted responses by domain and format. Among affected submissions, 55.1\% contained AI-supplied content for a subjective or person-specific response, such as attitudes, preferences, or evaluations; 52.2\% for knowledge or cognitive performance, such as factual recall, reading comprehension, or reasoning; and 11.9\% for creative production. Bounded formats, such as choices, ratings, or exact answers, appeared in 52.6\% of affected submissions and free-form responses in 57.5\%, with shares summing to more than 100\% because a submission could contain several affected exchanges. We next examined whether assistance was associated with payment rejection.

\begin{figure}[htbp]
\centering
\includegraphics[width=\linewidth]{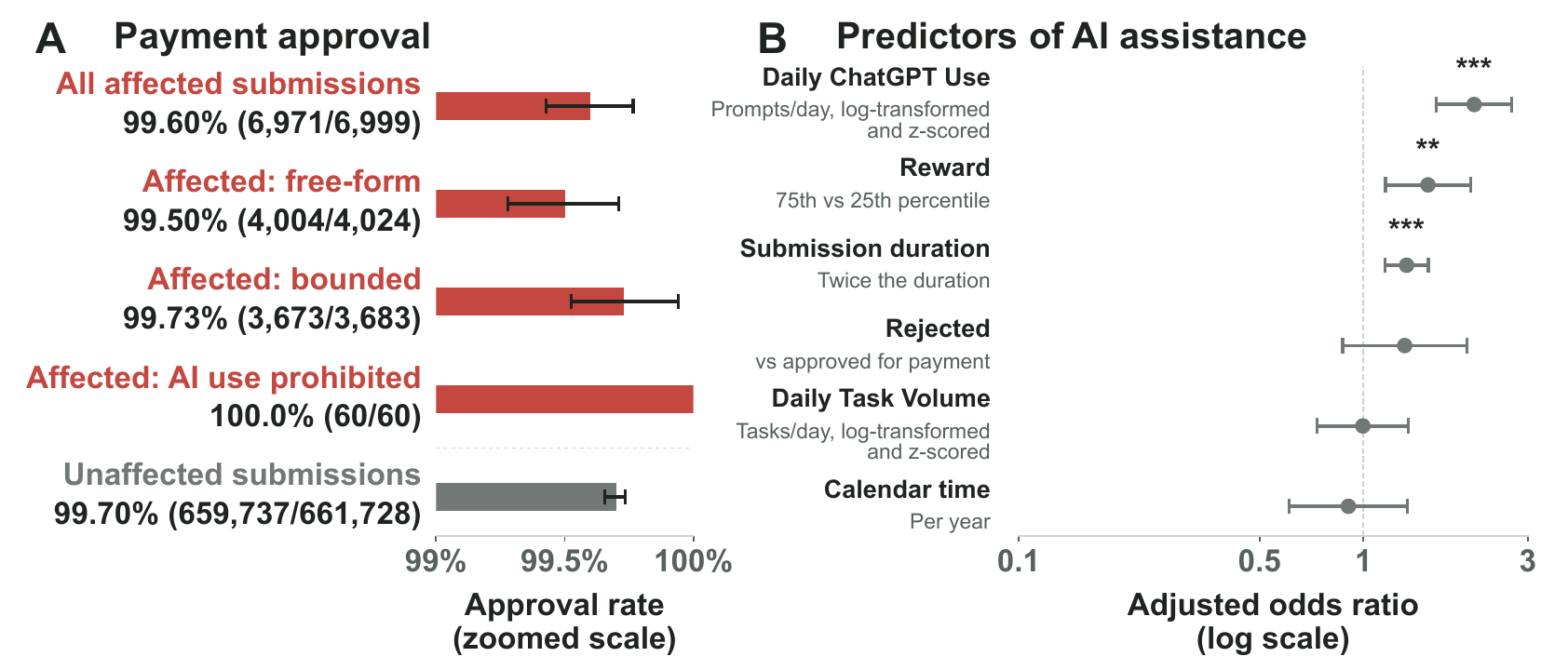}
\caption{\textbf{AI-assisted submissions were routinely approved, even when explicitly prohibited, and assistance was not associated with rejection.} Panels include only approved or rejected submissions.
(A) Approval rates for unassisted submissions and for assisted submissions overall, by response format, and with manually confirmed instructions prohibiting AI use (observed in the exchanges themselves). Approval rates exceeded 99\% across groups, including all submissions with confirmed prohibitions.
(B) Adjusted odds ratios, with 95\% confidence intervals, from a multiple logistic regression of submission-level assistance on the predictors shown. Assistance was associated with greater ChatGPT use, higher rewards, and longer submission duration, but not with rejection, submission volume, or calendar time. \(^{**}P < 0.01\), \(^{***}P < 0.001\).}
\label{fig:approval}
\end{figure}

\textbf{Workers were paid for virtually every submission on which they used AI assistance (Fig.~\ref{fig:approval}A).} Submissions with AI assistance were approved for payment 99.6\% of the time, as often for bounded responses (99.7\%) as for free-form ones (99.5\%). Even when observed study instructions explicitly prohibited AI use, all 60 submissions with a payment decision were approved and paid (100\%).

\textbf{AI assistance was not associated with rejection by researchers (Fig.~\ref{fig:approval}A and B).} Unaffected submissions were approved 99.70\% of the time, a rate statistically indistinguishable from affected ones. In a multiple logistic regression restricted to submissions approved or rejected for payment, rejection was not associated with assistance after adjusting for reward, duration, calendar time, and worker ChatGPT use and submission volume (P \textgreater{} .05; Fig.~\ref{fig:approval}B). Assistance also showed no time trend over the more than 3 year observation period and no association with workers' daily submission volume (both P \textgreater{} .05). Each standard-deviation increase in daily ChatGPT use, however, was associated with approximately twice the odds of submission-level assistance (OR, 2.09; 95\% CI, 1.63--2.69), and higher rewards (P = .003) and longer task duration (P \textless{} .001) were also associated with higher odds of assistance.

\section*{Discussion}

AI assistance was rare across the research activity we observed. Although 68\% of the workers we observed had used ChatGPT on a study, assistance appeared in only 1\% of submissions, with no detectable increase over more than three years. This contrasts with higher estimates from selected tasks, such as scientific text summarization \cite{ref2}, but the findings are compatible: assistance can be common in particular settings while remaining rare across workers' research activity overall. The breadth of study coverage and demographic comparability of our samples support the relevance of this finding to routine research participation on Prolific. By linking directly observed use to research submissions, we provide empirical evidence for a debate that has lacked direct estimates of how prevalent AI assistance actually is \cite{ref10}.

Our findings do not support the view that AI assistance currently poses an existential threat to online research. Instead, low observed prevalence and temporally localized use suggest limited displacement of human input. Importantly, the consequences of AI assistance are likely to depend on the task and outcome being studied. Prior simulations illustrate this heterogeneity: one estimated attenuation of a treatment effect from 0.40 to 0.36 SD at 10\% assistance \cite{ref5}, whereas another projected inflation of 2.3 percentage points at 4.4\% assistance \cite{ref3}. Thus, AI assistance can introduce meaningful bias in particular settings, but its direction and magnitude are not uniform. Our observed prevalence across a wide mix of studies was less than a quarter of the rates assumed in these estimates, further suggesting that estimates from selected tasks should not be generalized to online research as a whole.

While AI assistance is currently rare, our findings indicate platforms are poorly prepared should it become a wider threat. Workers were virtually always paid even when study instructions explicitly prohibited AI use, providing little immediate financial deterrent to such use. The limited scope of current safeguards compounds this problem: half of assisted submissions involved bounded responses, yet Prolific's optional LLM assistance detector screens only open-ended questions \cite{refprolific}. We therefore recommend safeguards cover bounded responses as well as free text. Assistance was also highly concentrated, with 5\% of workers accounting for nearly two thirds of assisted submissions. We consequently recommend complementing response-level checks with worker-level vetting that aggregates quality signals across studies, an approach that has previously improved data quality \cite{ref11}. Addressing these gaps matters for preserving the integrity of online research and the value of platforms that supply human input.

Several limitations apply. Although 76.5\% of donors reported using another AI service more than once a month, we did not collect histories from those services and cannot quantify their use on Prolific (Appendix). Excluding these services may understate assistance, while selecting weekly ChatGPT users may raise the estimate relative to Prolific overall, leaving the net direction of bias uncertain. The observed exchanges suggest that workers intended to use ChatGPT\textquoteright{}s output in their responses, although we cannot verify whether or how they incorporated it into their final submissions. Study-specific AI policies were usually unknown, so we cannot establish how often assistance violated instructions. ChatGPT histories were donated months before Prolific records were requested in separate studies, and workers who declined to share the latter had similar daily ChatGPT use to those who agreed, providing some reassurance about selective avoidance. Finally, our findings concern a single platform. Prolific is among the highest-quality platforms for recruiting research participants \cite{ref9}, making it an important setting for assessing whether online research can continue to elicit predominantly human input. The priority now is to detect repeated use across response formats, focusing on the few workers who account for most assistance while the overall rate remains low.

\section*{Materials and Methods}

We randomized 2,500 workers to direct questioning or a double-list experiment about lifetime AI assistance, irrespective of permission (Appendix S1).
Of 767 weekly ChatGPT users who donated chat histories in an unrelated earlier study, 425 subsequently donated Prolific submission records, yielding 408 workers with eligible submissions (Appendix S2.1).
Eligible submissions lasted no more than two hours and fell entirely within observed ChatGPT coverage (Appendix S2.3--S2.4). GPT-5.6 Luna classified overlapping exchanges for task relevance, AI assistance, response domain and format, and visible AI policies. Validation against human annotations of 100 submissions yielded Cohen's \(\kappa = 0.76\) for submission-level assistance, with exchange-level \(\kappa = 0.88\) for response domain and \(0.82\) for format. We quantified concentration using Gini coefficients. Worker-level bootstrapping provided confidence intervals for submission-level prevalence. Adjusted logistic regression examined correlates of assistance among approved or rejected submissions (Appendix S2.7). The University of Pennsylvania IRB approved the study. Participants provided informed consent.

\section*{Acknowledgments}

We are grateful to Dan\mbox{-}Mircea Mirea, Dylan Thurgood, Juliette Zaccour and seminar participants at HPI for helpful feedback. This project was supported in part by Schmidt Sciences (N.K.R.S.), Penn MEDIATED (N.K.R.S. and S.C.G.), the Walton Family Foundation (D.F.), and the Pennsylvania Department of Health's Health Research Formula Funds (HRFF) (S.C.G.). Content is the authors' responsibility and does not necessarily reflect the funders' views.

\section*{Author Contributions}

N.K.R.S. and M.T. conceived the research; N.K.R.S. and D.F. collected the data; N.K.R.S. analyzed data; L.U. and S.C.G. supervised the research; N.K.R.S. and M.T. wrote the first draft; and D.F., L.U., and S.C.G. provided intellectual input and revised the paper.

\section*{Competing Interests}

The authors have no competing interests to declare.


\beginsupplement

\begin{center}
{\Large\bfseries Supplementary Information}\par\medskip
{\large AI-Assisted Submissions in Online Research Are Rare and Highly Concentrated but Routinely Approved}\par\medskip
Neil K. R. Sehgal, Manuel Tonneau, Dunigan Folk, Lyle Ungar, and Sharath Chandra Guntuku\par\medskip
{\large Extended Methods}
\end{center}
\section{Survey of AI assistance}

\subsection{Participants and design}
On July 28, 2026, we recruited 2,500 adult Prolific participants residing in the United States, United Kingdom, or Canada. We distributed the study to a standard availability sample without demographic quotas or an approval-rate requirement. We randomly assigned participants with equal probability to a direct-report cell or one of four double-list cells. The realized sample included 504 direct-report respondents and 1,996 double-list respondents (cell sizes: 502, 495, 504, and 495). Direct-report respondents answered the sensitive item as a yes-or-no question and did not complete the list experiment. Double-list respondents completed both lists but did not answer the sensitive item directly. The sensitive item read: ``I have used ChatGPT or another AI tool to help complete a Prolific study.'' Because this wording does not specify whether the use violated study instructions, it measures broad lifetime AI assistance rather than unauthorized use specifically.

\subsection{Comparison with the active Prolific pool}
We compared the survey sample with the active Prolific pool within each country: United States (601 survey participants; 101,954 active workers), United Kingdom (1,779; 53,504), and Canada (120; 12,785). Available demographic screeners covered age, sex, employment status, ethnicity, first language, education, student status, country of birth, and nationality. Differences were modest on most demographic screeners, with larger imbalances in previous approved submissions. Workers with fewer than 100 prior approved submissions were underrepresented in all three countries.

\subsection{Control-item selection}
The control items were selected using two pilots. Candidate items were evaluated using their joint response distributions, floor and ceiling frequencies, ambiguity, cross-pilot stability, and within-list associations. Each final list contained a high-, medium-, and low-endorsement item.

\textbf{List A}

\begin{enumerate}
\def\labelenumi{\arabic{enumi}.}
\item
  ``In the past 30 days, I started a Prolific study but decided not to submit it.''
\item
  ``In the past 30 days, I completed a Prolific study that required me to join at a scheduled time.''
\item
  ``The most recent Prolific study I submitted before this survey asked me to make choices about money or payoffs (for example, choosing between gambles or deciding how to divide money).''
\end{enumerate}

\textbf{List B}

\begin{enumerate}
\def\labelenumi{\arabic{enumi}.}
\item
  ``The most recent Prolific study I submitted before this survey asked mainly about opinions, preferences, or attitudes.''
\item
  ``The most recent Prolific study I submitted before this survey required me to listen to audio or watch a video.''
\item
  ``The most recent Prolific study I submitted before this survey asked about politics, government, or current events.''
\end{enumerate}

\subsection{Double-list procedure}
Double-list respondents were asked to report how many statements were true without identifying which statements were true. They first had to complete a worked count example correctly. Respondents who initially answered incorrectly reviewed the instructions and repeated the example before proceeding.

The four cells crossed list order with the list containing the sensitive item:

\begin{enumerate}
\def\labelenumi{\arabic{enumi}.}
\item
  List A plus sensitive item, followed by List B;
\item
  List A, followed by List B plus sensitive item;
\item
  List B plus sensitive item, followed by List A; or
\item
  List B, followed by List A plus sensitive item.
\end{enumerate}

Thus, every double-list respondent answered both lists and encountered the sensitive item exactly once. Statement order was randomized within each list. Response options ranged from 0 to 3 for control lists and from 0 to 4 for lists containing the sensitive item.

\subsection{Statistical analysis}
For each list, prevalence was estimated as the difference between the mean count when the list contained the sensitive item and the mean count when it did not. The primary indirect estimate was the equally weighted average of the List A and List B estimates.

We calculated percentile 95\% confidence intervals using 10,000 respondent-level bootstrap samples stratified by the five randomized cells. Each double-list respondent's two responses remained together during resampling. The direct--indirect difference was calculated within the same bootstrap samples. We also examined endpoint frequencies, Blair--Imai design-effect tests, list-order interactions, and demographic balance. No respondents were excluded based on duration or comprehension attempts.

Endpoint frequencies were low. Counts at the maximum were observed for 5.1\% and 4.4\% of the List A and List B control conditions, respectively, and counts of four occurred for 1.1\% and 1.0\% of their corresponding sensitive-item conditions. Blair--Imai tests found no evidence that adding the sensitive item changed responses to the control items (List A, (P=1.000); List B, (P=0.318)). The average second-versus-first list-order interaction was 6.7 percentage points but was imprecisely estimated (95\% CI, \(-\)3.6 to 17.2; (P=0.213)). Demographic balance tests detected no differences after Holm adjustment.

\section{Linking ChatGPT histories to Prolific records}

\subsection{Participants and design}
For the analysis of linked histories, we recontacted participants from a separate study of donated ChatGPT histories conducted in winter 2026. The recontact group consisted of 767 participants. The earlier study recruited adults residing in the United States, United Kingdom, or Canada who had completed more than 10 Prolific studies and had indicated ChatGPT use on Prolific's AI screener. After consent, eligibility additionally required at least 1 month of ChatGPT account use and ChatGPT use at least once per week in a typical month. Among 4,600 entrants who answered both study eligibility questions, 2,985 (64.9\%) met the weekly-use criterion and 2,884 (62.7\%) met both the weekly-use and account-tenure criteria. These criteria intentionally selected established ChatGPT users and experienced Prolific workers. Of these 2,884 participants, 767 completed the full study and donated their ChatGPT history, forming the recontact frame.

From July 23 through August 23, 2026, members of the recontact frame were invited through Prolific to upload their Prolific submission-history CSV. The invitation did not state that recipients had been selected because they had previously donated a ChatGPT history, and the follow-up consent form did not specifically name linkage to those histories. It stated that the researchers ``may also use a coded identifier to connect your responses to prior research records.'' The full protocol was approved by the University of Pennsylvania IRB, and participants provided informed consent.

We received 425 ChatGPT histories, corresponding to 55.4\% of the 767 person recontact frame. Of these participants, 408 contributed at least 1 primary eligible task window. Our observational analysis of linked histories therefore concerns a multiply selected cohort: prior ChatGPT donors who returned for follow-up, uploaded a valid Prolific history, and contributed at least 1 task during their observed ChatGPT period.

\subsection{Sample comparisons and selection into donation}
We compared the 425 participants with accepted Prolific-history uploads against the other 342 members of the 767-person recontact frame using ChatGPT measures and demographics collected before follow-up. Median conversations per observed ChatGPT-history day were 0.252 among donors and 0.259 among non-donors (Cliff's \(\delta\), -0.009; P = .835). Median user turns per observed day were 1.18 and 1.18, respectively (\(\delta\), -0.002; P = .970). Self-reported ChatGPT-use frequency distributions were also similar (Cramér's V, 0.047; P = .793). Donors more often reported more than 12 months of ChatGPT use (271/425, 63.8\%) than non-donors (201/342, 58.8\%). However, the tenure-distribution comparison did not remain significant after Holm adjustment across the 9 ChatGPT-use tests.

Demographic differences between donors and non-donors were also minimal after adjustment: age group (V, 0.098; adjusted P = .274), gender (V, 0.094; adjusted P = .138), race or ethnicity (V, 0.092; adjusted P = .274), and country (V, 0.051; adjusted P = .369). Therefore, within this already selected cohort of established ChatGPT users, follow-up donation was not materially associated with greater observed ChatGPT-use intensity or large demographic differences.

The 408 eligible donors differed from 2,500 survey participants most clearly in country composition (Cramér's V = 0.228; Holm-adjusted P \textless{} .001). The survey sample included more participants from the UK (71.2\% vs 40.4\%), whereas the eligible donors included more from the US (24.0\% vs 47.8\%). This was partly due to the survey going live during UK morning hours. Age differences were smaller (V, 0.112; adjusted P \textless.001), including fewer participants aged 55 years or older (17.4\% vs 7.6\%). Gender or sex distributions differed modestly (V, 0.052; adjusted P = .039); race or ethnicity was not detectably different after adjustment (V, 0.047; adjusted P = .089).

To compare survey participants with donors on ChatGPT use, we randomly selected 650 of the original 2,500 participants for invitation to a follow-up survey in September 2026, targeting 500 completions. Participants answered the account-use duration and typical-month ChatGPT-use frequency questions used to screen the original ChatGPT donors. Among the 500 respondents, 63.4\% reported at-least-weekly ChatGPT use, with 32.2 \% reporting at-least-daily use.

\paragraph{Use of other AI services.}
Participants were asked, ``Aside from ChatGPT, do you use any of these other AI chatbots more than once a month? Please select all that apply.'' Among the 408 eligible donors, 312 (76.5\%) selected or named another chatbot or AI service including Gemini, Claude, Character.AI, Snapchat MyAI, and other services. We did not collect histories from these services.

\subsection{Prolific records and primary task windows}
We call the interval from a Prolific task record's start timestamp through its completion timestamp a task window. We then restricted to tasks no more than 2 hours. This 2-hour cap reduced the influence of stale or abandoned sessions and chance overlap; 1-hour, 4-hour, and uncapped windows were examined in sensitivity analyses. A primary eligible task window also had to fall entirely between that participant's first and last visible timestamped ChatGPT user messages. This restriction avoided classifying tasks outside the observed ChatGPT-history period as observed nonuse. This yielded 712,930 task windows from 408 workers. Eligibility did not depend on approval status of the task. Approved, returned, screened-out, rejected, timed-out, and other recorded task statuses could contribute. However, for certain analyses we restrict to only approved or rejected submissions. The eligible task windows spanned December 2022 through February 2026. Prolific submission histories do not include unique study identifiers, so we approximated the number of distinct studies by counting unique publisher-name and study-title combinations. Separate studies conducted under the same publisher name and title would be counted together, potentially underestimating the number of distinct studies.

\subsection{Temporal overlap and exchange construction}
A temporal overlap occurred when at least 1 visible timestamped ChatGPT user message fell between task start and completion, inclusive. This does not mean that the full ChatGPT conversation occurred during the task. This approach identified 12,564/712,930 (1.76\%) task windows among 325/408 (79.7\%) workers.

For annotation, an exchange began with 1 in-window user message and included all subsequent in-window assistant messages before the next user message in the same ChatGPT conversation. One task could contain several exchanges and could overlap several ChatGPT conversations. The 12,564 overlapping tasks contained 46,691 target exchanges.

\subsection{Annotation}
Each API request represented 1 overlapping task and asked the model to label every target exchange. Inputs contained the Prolific study title and publisher, the number of concurrently active eligible Prolific windows, all visible user and assistant messages inside the window, up to 4 immediately preceding messages per ChatGPT conversation marked as prior context, stable message and exchange identifiers, minutes relative to task start, and the model slug when available. GPT-5.6 Luna processed requests using medium reasoning, no tools, and strict structured output.

The classifier separately assessed task relationship, whether the user supplied an actionable request or task item, what contribution the assistant delivered, the supported participant-response type, response format, visible AI-policy context, and observable evidence flags. At task level, visible AI policy was classified as required or encouraged, prohibited, or not visible. Prohibition required explicit study-specific instructions; absence of a visible policy did not establish permission. The annotation codebook was iteratively developed through manual review of 200 overlapping exchanges.

Some detailed labels were reduced deterministically after annotation. Current-task interaction was positive only for current\_\allowbreak{}task. Usable task help combined model\_\allowbreak{}added\_\allowbreak{}measured\_\allowbreak{}response and usable\_\allowbreak{}help\_\allowbreak{}without\_\allowbreak{}model\_\allowbreak{}added\_\allowbreak{}response. The 4 participant-response types combined personal/autobiographical, attitude/preference/belief, and evaluation/annotation into ``Personal, attitudinal, or evaluative''; knowledge/\allowbreak{}factual recall/\allowbreak{}forecast/\allowbreak{}translation, reading comprehension, and reasoning/problem solving into ``Knowledge, comprehension, or reasoning''; and retained ``Creative production'' and ``Other or unclear.'' Response format used the 3 categories specified directly in the prompt: ``Prespecified or bounded response,'' ``Free-form response,'' and ``Structured, mixed, or unclear.'' Translation was classified as knowledge and, when the answer required translated prose rather than an objectively bounded token, as free-form.

Task-level endpoints used an any-positive rule. A task was positive if at least 1 constituent exchange was positive. Participant-response types and formats were nonexclusive at task level because different exchanges within the same task could carry different labels.

\subsection{Annotation Validation}
The validation sample was a random sample of 100 overlapping tasks, containing 362 target exchanges from 58 workers. One human annotator, the lead author, coded all exchanges. At exchange level, raw agreement and Cohen's \(\kappa\) were 0.942 and 0.844 for current-task interaction, 0.925 and 0.835 for an actionable request or item, 0.876 and 0.708 for usable task help, and 0.925 and 0.840 for ChatGPT-contributed responses. After any-positive aggregation across exchanges, task-level raw agreement and \(\kappa\) were 0.870 and 0.715 for current-task interaction, 0.820 and 0.626 for usable task help, and 0.880 and 0.759 for ChatGPT-contributed responses. Conditional on both raters assigning an applicable downstream taxonomy, \(\kappa\) was 0.875 across the 4 participant-response types (214 exchanges) and 0.821 across the 3 response formats (218 exchanges).

\paragraph{Validation of explicit AI prohibitions.}
The original validation sample contained too few observed AI prohibitions to assess this label reliably. We therefore manually reviewed all 73 assisted submissions flagged as explicitly prohibiting AI use by the Luna classifier. Prohibitions were confirmed in 61 submissions (precision, 83.6\%). Recall cannot be estimated, and this selected subset cannot estimate the prevalence of prohibited assistance. Figure 2 reports payment approval among the 60 submissions recorded as approved or rejected for payment; one returned submission was excluded.

\subsection{Statistical analysis}
Primary prevalence estimates were task weighted. Confidence intervals used 4,000 worker-level bootstrap samples. Worker-history estimates indicate whether a worker reached an endpoint at least once across the observed history.

Worker concentration was summarized using Lorenz curves and Gini coefficients. Confidence intervals used the same 4,000 worker-level resamples. Within-task timing was calculated as elapsed time from task start divided by total task duration. An exchange comprised the user anchor and its following assistant response sequence.

A multiple logistic regression modeled the ChatGPT-contributed-response endpoint across primary eligible task windows. We restricted this analysis to tasks that were approved or rejected for payment, and excluded those that were returned, timed out, or screened out. Predictors were within-currency reward percentile, log\textsubscript{2} task duration, continuous calendar year, task approval status (approved vs. rejected), and 2 worker-level measures. Worker ChatGPT-use intensity was the number of visible timestamped user prompts divided by days from the first to last visible ChatGPT message. Worker eligible-task density was the number of primary eligible task windows divided by the same observed-history days. Both rates were transformed as ln(1 + rate) and standardized to mean 0 and SD 1 across 408 workers.

\section{Annotation Prompt}

We annotated exchanges using OpenAI's gpt-5.6-luna model with medium reasoning effort in August 2026. Each request contained one task's transcript and contextual information and returned exchange-level labels under a strict JSON schema. No tools were supplied. Temperature, nucleus-sampling probability, and random seed were not explicitly specified and left at the defaults. The maximum output-token limit was 128,000 per request.

\subsection{System prompt}
\begin{Verbatim}
You are a research classifier analyzing donated ChatGPT messages that overlap in time with one Prolific research task.

Label every TARGET EXCHANGE separately. A target exchange begins with one USER message inside the task window and includes every following ASSISTANT message before the next user message in that ChatGPT conversation. Messages marked PRIOR CONTEXT may clarify a target exchange but must not receive their own label.

Use only visible evidence. Temporal overlap alone does not establish task relationship. The study title and publisher provide context but cannot establish relationship by themselves when the exchange is generic. Do not infer that model output was copied, submitted, authorized, or prohibited. Do not call behavior cheating.

Make five exchange-level judgments and one task-level AI-policy judgment:

1. TASK RELATIONSHIP
- current_task: concretely concerns this task's content, instructions, workflow, or response.
- not_current_task: clearly unrelated to this task or concerns only general Prolific administration, payment, access, or account help.
- indeterminate: visible evidence cannot reliably establish or rule out a relationship.

2. TASK RESPONSE OPPORTUNITY
Judge only the target USER message and prior context, independently of the assistant's response.
- actionable_request_or_item: an explicit request/command or an actual task item ChatGPT could act on. A bare survey question, rating item, answer choices, or screenshot can be an implicit request.
- no_actionable_request_context_only: only instructions, consent language, browser requirements, stimulus material, a scenario, or other context is supplied without an answerable item or request.
- not_assessed_relationship_not_established: required whenever task_relationship is not_current_task or indeterminate.
The assistant volunteering a summary cannot turn context-only user input into an actionable request.

3. DELIVERED CONTRIBUTION
Judge visible assistant output, not merely user intent.
- model_added_measured_response: the assistant supplied or materially added content that could answer any part of what the study sought from the participant. This includes personal narratives, attitudes, forecasts, choices, ratings, summaries requested as task outputs, factual answers, calculations, reasoning, classifications, evaluations, and creative responses. A partial usable answer counts. If the participant supplied a scaffold but the model materially added measured content, use this label.
- usable_help_without_model_added_response: the assistant gave usable procedural or interpretive help, or edited/translated/expressed a response whose central substance the participant had already supplied, without materially adding what the study sought to measure.
- no_observed_usable_current_task_help: no visible response, refusal, irrelevant response, request for missing information, mere acknowledgement/restatement, or otherwise no help usable for the current task.
- not_assessed_relationship_not_established: required whenever task_relationship is not_current_task or indeterminate.

4. RESPONSE TARGET
Characterize the primary response the assistant supported. Use personal_autobiographical; attitude_preference_belief; evaluation_annotation; knowledge_factual_recall_forecast; reading_comprehension_interpretation; reasoning_problem_solving; creative_production; other_unclear; or not_applicable_no_supported_response.
Group beliefs with attitudes and preferences. Group forecasts, predictions, expectations, factual estimates, and translation or bilingual language rendering with knowledge and factual recall. Translation must always be labeled knowledge_factual_recall_forecast, not reading_comprehension_interpretation, even when the source text is supplied in the task. Distinguish those from reasoning, and distinguish both from other comprehension of material supplied in the task. Use not_applicable_no_supported_response for purely procedural help or no usable help.

5. RESPONSE FORMAT
Classify the response space, not the observed answer length or whether the answer is objectively correct.
- prespecified_or_bounded_response: the task requires an answer within a response space specified before the respondent answers. This includes a choice, category, rating, yes/no response, number, date, named entity, exact word, code, or calculated result.
- free_form_response: the respondent determines substantive wording or content and multiple meaningfully different responses could satisfy the task. A word or sentence limit does not make a response bounded. Translation or bilingual language rendering is always free-form unless the task requires selecting from prespecified answer choices. This remains true when a translation has a correct meaning, asks who or what did something, or could be answered with one short phrase. Summarization, explanation, personal narrative, and other brief written responses are also free-form unless the task presents fixed choices.
- structured_mixed_or_unclear: code, tables, structured artifacts, participant-generated media, multiple central response formats, or insufficient evidence to distinguish bounded from free-form.
- not_applicable_no_supported_response: no measured response was supported.

Decision rule: could the expected response space or value type be specified before seeing the respondent's wording? If yes, use prespecified_or_bounded_response. If no, use free_form_response. If both or unclear, use structured_mixed_or_unclear.

Translation priority rule: when the supported response translates or renders language bilingually, assign response_target=knowledge_factual_recall_forecast and response_format=free_form_response. Do not classify a translation as prespecified_or_bounded_response merely because it has a correct semantic answer, a short answer, or a limited set of people or objects mentioned in the source. Override free_form_response only when visible task instructions require choosing from prespecified response options rather than producing translated wording.

At the task level, assign exactly one visible_ai_policy label using the study title and all supplied messages. Use required_or_encouraged only for an explicit task-specific instruction inviting or requiring AI use; prohibited only for an explicit task-specific prohibition; and no_visible_ai_policy when neither is visible. An AI-themed study title does not by itself establish a policy. A title explicitly instructing the participant to use ChatGPT can establish required_or_encouraged. If both encouragement and prohibition appear, classify the policy that applies to the observed external ChatGPT use; if that cannot be distinguished, use prohibited because a task-specific prohibition is visible.

Evidence message IDs should be the smallest supplied set supporting the labels.

Consistency rules:
- If task_relationship is not_current_task or indeterminate, set task_response_opportunity and delivered_contribution to not_assessed_relationship_not_established, and response_target and response_format to not_applicable_no_supported_response.
- If task_response_opportunity is no_actionable_request_context_only, delivered_contribution cannot be model_added_measured_response, and response_target and response_format must be not_applicable_no_supported_response. The assistant may still have delivered usable procedural or interpretive help, so distinguish that from no_observed_usable_current_task_help.
- If delivered_contribution is no_observed_usable_current_task_help, response_target and response_format must be not_applicable_no_supported_response.
- If delivered_contribution is usable_help_without_model_added_response, use a substantive response target/format only when the assistant supports a participant-originated measured response; otherwise use not_applicable_no_supported_response.
- For every translation or bilingual language-rendering response without prespecified answer choices, response_target must be knowledge_factual_recall_forecast and response_format must be free_form_response.
- Every exchange ID and evidence message ID must exactly match an ID supplied in the input.

Return only JSON matching the supplied strict schema.

USER TEMPLATE

Classify the target exchanges in this Prolific task window.

TASK ID:
{task_id}

PROLIFIC STUDY TITLE:
{study_title}

PUBLISHER:
{publisher}

CONCURRENT OTHER PROLIFIC TASK WINDOWS:
{concurrent_task_window_count}

TARGET EXCHANGE IDS IN REQUIRED OUTPUT ORDER:
{exchange_ids}

TRANSCRIPT:
{transcript}
\end{Verbatim}
\subsection{Output schema}
\begin{Verbatim}
TASK_RELATIONSHIP = [
    "current_task",
    "not_current_task",
    "indeterminate",
]
TASK_RESPONSE_OPPORTUNITY = [
    "actionable_request_or_item",
    "no_actionable_request_context_only",
    "not_assessed_relationship_not_established",
]
DELIVERED_CONTRIBUTION = [
    "model_added_measured_response",
    "usable_help_without_model_added_response",
    "no_observed_usable_current_task_help",
    "not_assessed_relationship_not_established",
]
RESPONSE_TARGET = [
    "personal_autobiographical",
    "attitude_preference_belief",
    "evaluation_annotation",
    "knowledge_factual_recall_forecast",
    "reading_comprehension_interpretation",
    "reasoning_problem_solving",
    "creative_production",
    "other_unclear",
    "not_applicable_no_supported_response",
]
RESPONSE_FORMAT = [
    "prespecified_or_bounded_response",
    "free_form_response",
    "structured_mixed_or_unclear",
    "not_applicable_no_supported_response",
]
VISIBLE_AI_POLICY = [
    "required_or_encouraged",
    "prohibited",
    "no_visible_ai_policy",
]
HUMAN_FIELD_OPTIONS: dict[str, list[dict[str, str]]] = {
    "task_relationship": [
        {
            "value": "current_task",
            "label": "Current Prolific task",
            "description": "The exchange concretely concerns the content, instructions, workflow, or response for the currently overlapping task.",
        },
        {
            "value": "not_current_task",
            "label": "Clearly not the current task",
            "description": "The exchange is unrelated to the current task or concerns general Prolific administration, payment, access, or account help.",
        },
        {
            "value": "indeterminate",
            "label": "Indeterminate",
            "description": "The visible title, transcript, and timing do not establish or rule out a relationship reliably.",
        },
    ],
    "task_response_opportunity": [
        {
            "value": "actionable_request_or_item",
            "label": "Actionable request or task item",
            "description": "The user supplied a request, command, or actual item ChatGPT could act on. A bare survey item or screenshot can count implicitly.",
        },
        {
            "value": "no_actionable_request_context_only",
            "label": "No actionable request: context only",
            "description": "The user supplied only instructions, consent text, browser requirements, stimulus material, or a scenario without an answerable item or request.",
        },
        {
            "value": "not_assessed_relationship_not_established",
            "label": "Not assessed: task relationship not established",
            "description": "Automatically used when relationship is clearly not current or indeterminate.",
        },
    ],
    "delivered_contribution": [
        {
            "value": "model_added_measured_response",
            "label": "Model added measured-response content",
            "description": "The assistant supplied or materially added content that could answer any part of what the study sought from the participant.",
        },
        {
            "value": "usable_help_without_model_added_response",
            "label": "Usable help without model-added response",
            "description": "The assistant provided procedural or interpretive help, or edited/expressed a response whose core substance the participant had already supplied.",
        },
        {
            "value": "no_observed_usable_current_task_help",
            "label": "No observed usable current-task help",
            "description": "No response, refusal, irrelevant reply, mere restatement, or otherwise no visible help usable for the current task.",
        },
        {
            "value": "not_assessed_relationship_not_established",
            "label": "Not assessed: task relationship not established",
            "description": "Automatically used when relationship is clearly not current or indeterminate.",
        },
    ],
    "response_target": [
        {
            "value": "personal_autobiographical",
            "label": "Personal or autobiographical",
            "description": "Personal facts, attributes, experiences, feelings, or self-report.",
        },
        {
            "value": "attitude_preference_belief",
            "label": "Attitude, preference, or belief",
            "description": "Subjective attitudes, preferences, values, or beliefs about oneself or the world.",
        },
        {
            "value": "evaluation_annotation",
            "label": "Evaluation or annotation",
            "description": "Rating, classifying, coding, comparing, or judging supplied objects or responses.",
        },
        {
            "value": "knowledge_factual_recall_forecast",
            "label": "Knowledge, factual recall, or forecast",
            "description": "Factual knowledge or recall, translation, factual estimates, expectations, forecasts, or predictions, rather than inference from supplied material or a multistep solution.",
        },
        {
            "value": "reading_comprehension_interpretation",
            "label": "Reading comprehension or interpretation",
            "description": "Understanding, summarizing, recalling, or interpreting material supplied in the task.",
        },
        {
            "value": "reasoning_problem_solving",
            "label": "Reasoning or problem solving",
            "description": "Mathematics, logic, puzzles, calculations, or objectively constrained multistep reasoning.",
        },
        {
            "value": "creative_production",
            "label": "Creative production",
            "description": "Original writing, naming, idea generation, images, or other creative output.",
        },
        {
            "value": "other_unclear",
            "label": "Other or unclear",
            "description": "A supported response is visible, but no response-target category fits clearly.",
        },
        {
            "value": "not_applicable_no_supported_response",
            "label": "Not applicable: no supported task response",
            "description": "Use when the contribution does not support a measured task response, such as procedural instruction help, or when no usable help was delivered.",
        },
    ],
    "response_format": [
        {
            "value": "prespecified_or_bounded_response",
            "label": "Prespecified or bounded response",
            "description": "The answer must use a response space specified before the response, such as a choice, category, rating, yes/no value, number, date, named entity, exact word, code, or calculated result.",
        },
        {
            "value": "free_form_response",
            "label": "Free-form response",
            "description": "The respondent determines substantive wording or content. A word limit, sentence limit, translation, summary, or request to keep an answer short remains free-form.",
        },
        {
            "value": "structured_mixed_or_unclear",
            "label": "Structured, mixed, or unclear",
            "description": "Code, tables, structured artifacts, participant-generated media, multiple central formats, or a format that cannot be determined.",
        },
        {
            "value": "not_applicable_no_supported_response",
            "label": "Not applicable: no supported task response",
            "description": "Use when the assistant did not support a measured task response.",
        },
    ],
    "visible_ai_policy": [
        {
            "value": "required_or_encouraged",
            "label": "AI use required or encouraged",
            "description": "Visible task-specific instructions require, invite, or encourage the participant to use ChatGPT or another AI system.",
        },
        {
            "value": "prohibited",
            "label": "AI use prohibited",
            "description": "Visible task-specific instructions prohibit ChatGPT, generative AI, or external AI assistance.",
        },
        {
            "value": "no_visible_ai_policy",
            "label": "No visible AI policy",
            "description": "No task-specific AI permission, encouragement, requirement, or prohibition is visible in the supplied title and messages.",
        },
    ],
}
\end{Verbatim}

\end{document}